\documentstyle[12pt,graphics]{article}  
  
\newcommand{\news}{\setcounter{equation}{0}}

\def\eqn{\begin{equation}}  
\def\eeqn{\end{equation}} 
\def\arr{\begin{array}}  
\def\earr{\end{array}} 
\def\eqna{\begin{eqnarray}}  
\def\eeqna{\end{eqnarray}}  
\def\a{\alpha} 
\def\b{\beta}  
 
\def\s{\sigma}  
\def\d{\delta}

\def\O{\Omega}  
\def\e{\epsilon} 
\def\th{\theta}  
\def\m{\mu}  
\def\n{\nu}  
\def\la{\lambda}  
 
\def\t{\tau}  
\def\p{\partial}  
\def\g{\gamma}  
\font\mybb=msbm10 at 12pt  
\def\bb#1{\hbox{\mybb#1}}

\def\bE {\bb{E}}

\def\bv{{\bf v}} 
\def\bx{{\bf x}} 
 
\def\bY{{\bf Y}} 
\def\vdot{\dot{\bf v}}  
\def\bn{{\bf n}}  
  
\def\bk{{\bf k}}  
 
\def\bS{{\bf S}} 
 
\begin{document}  
  
\begin{titlepage}  
  
\begin{flushright}  
AEI-1999-16\\  
PUPT-1843  
\end{flushright}  
  
\vspace{1.5cm}  
  
\begin{center}  
  
{\Large \bf{Radiation from Accelerated Branes}}  
  
\vspace{1.8cm}  
  
Mohab Abou-Zeid$^1$\footnote{abouzeid@aei-potsdam.mpg.de} 
and Miguel S.\ Costa$^2$\footnote{miguel@feynman.princeton.edu}  
  
\vspace{.5cm}  
  
$^1${\em Max Planck Institut f\"{u}r Gravitationsphysik, Albert Einstein 
Institut,\\Am  
M\"{u}hlenberg 1, D-14476 Golm, Germany}  
  
\vspace{.5cm}  
  
$^2${\em Joseph Henry Laboratories, Princeton University, \\ 
Princeton, New Jersey 08544, USA}  
  
\vspace{1.5cm}  
  
\begin{abstract}  
The radiation emitted by accelerated fundamental strings and D-branes is 
studied within 
the linear approximation to the supergravity limit of string theory. We show 
that scalar,
gauge field and gravitational radiation is generically emitted by such branes. 
In the case where an external scalar field accelerates the branes,
we derive a Larmor-type formula for the emitted scalar radiation 
and study the angular distribution of the outgoing energy flux. The 
classical radii of the branes are calculated by means of the corresponding 
Thompson 
scattering cross sections.  Within the linear approximation, the interaction 
of the  external scalar field  with the 
 velocity fields of the  branes  gives a contribution to the observed gauge 
field and gravitational radiation.
\end{abstract}  

\end{center}   

\end{titlepage}  
  
\section{Introduction}  
  
In addition to fundamental strings, nonperturbative superstring theory 
contains various types of extended BPS objects which play a key role 
in the web of dualities that have been uncovered in recent years, 
notably NS solitons~\cite{CHS,PKT,CMH} and  D-branes~\cite{JP1,JPb}, which 
carry R-R charges.  The theory also contains non-BPS solitonic states, 
some of which are stable~\cite{ASn}.  
  
Among the BPS states, the NS 5-branes  
appear to be rather complicated objects whose dynamics is poorly 
understood, mainly because no tractable CFT description is 
known for them. However, it has been argued that the dynamics of NS 5-branes 
is 
given by a new theory of self-dual closed strings in six dimensions.  
This formulation has proved remarkably fruitful in studying the BPS 
spectrum of compactified string and M theories~\cite{DVV}, even though 
the nature of the posited six-dimensional string theory without 
gravity remains somewhat mysterious.  
  
By contrast, D-branes admit a very simple CFT description, at least at 
weak coupling, as  hypersurfaces where open strings can end~\cite{JP1}. This 
has made 
it possible to use world-sheet techniques to study various aspects of 
the dynamics of non-relativistic as well as relativistic D-branes, and 
to perform explicit computations of scattering amplitudes involving 
D-branes. This includes processes in which fundamental closed strings 
scatter off a D-brane~\cite{KT,JB,GHKM,GM2,HK}, as well as processes in which 
a D-brane 
scatter off an identical D-brane~\cite{Bachas}.  Furthermore, D-branes have 
a number of properties which suggest that they are small at weak 
string coupling $g_s$~\cite{KT,SHS2}: in string units, their 
R-R charge is proportional to $g_s$ which is small and their tension 
is proportional to $1/g_s$ which is small compared to the tension 
of the NS solitons (proportional to $1/g_s^2$). This 
suggests that, in scattering processes in which a lower-dimensional 
D-brane scatter off a higher-dimensional D-brane, the former can be 
viewed as a probe of sub-stringy distances in the background geometry 
of the latter, and much can be learned about the physics at distances 
shorter than the ten-dimensional string and even the Planck lengths 
from this viewpoint~\cite{DKPS,DPS}.  In this framework, there are two 
descriptions of the interactions between D-branes: the supergravity 
interaction mediated by massless closed strings, and the probe quantum 
field theory which gives the effective dynamics of the various open 
string sectors.  The range of validity of the two descriptions are 
different: supergravity is reliable at distances large compared to the 
string scale, while sub-stringy effects are governed by the light open 
strings.  
   
However, most studies of string and D-brane scattering generally neglect  
a number of dynamical effects such as D-brane recoil and acceleration 
effects~\cite{PT,FPR,PU}. The latter are also important in the studies of the  
structure of black holes using strings and D-branes as probes~\cite{DPS}.  
For example, if the probe is excited, or the black hole is non-extremal,  
then supersymmetry is broken and the long range forces on an originally  
static probe do not exactly cancel. The resulting probe acceleration is 
$a \sim GM$  
where $G$ is the Newton constant and $M$ the mass of the black hole.  
For a D-brane background, this gives an acceleration $a\sim g_s$, which is 
small  
at weak string coupling. For an F-string, this gives $a\sim g_s^2$, which is  
even smaller. For the NS 5-brane, the acceleration is of order one. 
 
From the above considerations, it is clear that a better understanding of 
brane 
acceleration effects, which so far seem to have  
attracted little attention, is required. The aim of this paper is to start 
exploring 
these effects and to compute the radiation produced by moving branes. This 
may turn out to be of import to the 
understanding of string and M theory dynamics. For example, it was argued  
in~\cite{FS} that rigid fundamental strings can only be accelerated up to a  
certain value of the acceleration $a_c$ which is of the same order as, and  
possibly related to, the Hagedorn temperature $T_H$ of string theory. Beyond 
this  
value, the accelerated string appears to be unstable. It would be interesting  
to investigate whether this observation extends to dynamical strings and to 
other  
objects in string theory. In particular, this might suggest a maximal 
acceleration  
for the open strings tethered on a D-brane\footnote{We would like to thank 
Michael Green  
for suggesting this possibility.} or the self-dual closed strings making up 
the world-volume of a NS 5-brane. Recall that the Born-Infeld-type effective 
action for D-branes already implies the existence of a limiting world-volume 
electric field $E_{crit} = (2\pi \alpha ' c)^{-1}$, which is in dual 
correspondence with the limiting space-time velocity $c$~\cite{Bachas}. 
  
In the absence of  a second quantized theory of strings and branes, it is not 
clear how  
acceleration effects should be described in (quantum) string theory. At any 
rate, several  
aspects of the dynamics of accelerated strings and branes can be studied in 
the  
classical supergravity description. In this paper we will focus on  radiation  
effects from moving branes in the linear approximation to the classical limit 
of string theory.  
 
The fundamental string solution couples to the metric, the NS-NS 
antisymmetric 2-form  
potential and the dilaton and so one expects an accelerated F-string to 
produce  
dilatonic, \lq electric' NS-NS antisymmetric tensor field  and gravitational 
waves.  
In addition to the 
metric and dilaton fields, D-branes couple to the R-R gauge fields of the 
type II  
superstrings, so accelerated D-branes can be expected to emit dilatonic, 
gravitational 
and R-R radiation. Likewise, we expect accelerated NS 5-branes to emit 
dilatonic,  
gravitational and \lq magnetic' antisymmetric tensor field   waves. The 
$(p,q)$ strings 
of the type IIB theory~\cite{EW,JHS} are expected to emit both \lq electric' 
NS-NS 
antisymmetric tensor field  radiation and 
R-R 2-form gauge field radiation in addition to the dilatonic and 
gravitational radiation.  
We will be able to confirm some of these  
expectations by explicit calculation of the supergravity fields produced by 
accelerated  
strings and D-branes at large distances in the weak field approximation.   
  
We will restrict ourselves to situations in which  the  accelerated string or 
brane  
can be thought of as a localised source propagating in a background which is 
a small  
deformation of flat space-time at weak string coupling. This is possible for 
F-strings  
and D-branes, but not for NS 5-branes. In the latter case the space-time 
metric involves a  
harmonic function $H =1+\m/r^2$ in the four-dimensional transverse space with  
$\m$ a constant of order one in string units, and hence space-time is not  
flat at small $g_s$.   
  
The organization of the paper is as follows. In section 2, we review the 
system of  
coupled supergravity-brane field equations which we wish to solve in the 
linear 
approximation. We shall consider a weak external scalar field whose effect is 
to
accelerate the branes. The solutions for the scalar fields, the $p$-form gauge 
field and the metric fluctuation are studied in sections 3, 4 and
5, respectively. In the case of  the scalar fields, we  derive 
the energy flux of the emitted radiation and study its angular distribution. 
We calculate the corresponding Thompson scattering cross-sections and obtain 
the classical radii of the 
F-string and D$p$-branes, in agreement with expectations
from tree level string diagrams. The  differential equations obeyed by the 
gauge field and metric fluctuation are discussed. These can in principle be 
solved for the radiation fields, although we have not been able to do so 
explicitly. This is because the external scalar field responsible for 
accelerating the branes interacts 
with the velocity fields (i.~e.\ those not depending on the brane 
acceleration), which  gives 
a contribution to the observed gauge and gravitational  radiation fields that 
is not 
negligible within our approximations. We make some comments about possible 
extensions 
of the present work in the final section.

\section{Coupled Supergravity-Brane Field Equations} 
\news 
 
The effective action for string theory in the presence of a $p$-brane in the 
long  
wave-length limit has the general form 
\begin{equation}  
\label{eq:sum}  
 S = S_{bulk} + S_{brane}\ ,  
\end{equation}  
where $S_{bulk}$ is the appropriate supergravity action and $S_{brane}$ is 
the  
effective action for the dynamics of the world-volume fields on the brane. In 
the  
following sections, we will consider the field equations  
derived from the bosonic part of  the effective action~(\ref{eq:sum}) which 
describe the  
long-range bosonic fields outside a fundamental string or a D$p$-brane 
source. At weak string  
coupling $g_s$, the corresponding background geometries are nearly flat and 
it is  
consistent to work in an approximation in which the branes are described by 
a sharply 
localised  source  
propagating in a flat background. Our strategy will be to study the 
linearised theory  
around flat Minkowski space-time and to solve for the propagating small 
fluctuations  
of the NS-NS and (for the D-branes) the R-R fields.  
 
Here we start by writing the bosonic bulk action for a
$4$-dimensional space-time obtained by compactification of  10-dimensional 
space-time 
on a rectangular 6-torus $T^6 = T^p \times T^{(6-p)}$: the spatial directions 
along 
the $p$-brane  are wrapped around $T^p$, while $(6-p)$ of the remaining 
transverse 
directions are  compactified on $T^{(6-p)}$. Then we derive the form of the 
bosonic 
particle actions arising from the compactified $p$-brane actions, and  obtain 
the field equations 
of the coupled system. For branes accelerated by an  external scalar field, 
we derive the linearised field equations.  

\subsection{Compactified Effective Action} 
 
For the fundamental string, the non-zero bulk fields correspond to the 
bosonic sector  
of the 3-form version of $D=10$, $N=1$ supergravity. The  string frame 
action is  
\eqn 
\label{2.2}  
S_{bulk} = \frac{1}{2\kappa^2_{10}} \int d^{10} x \sqrt{-G}~e^{-2\phi_{10}} 
\left( R +4(\partial_a \phi_{10})^2 - \frac{1}{2\cdot 3!} {\cal H}^2 
\right)\ ,  
\eeqn  
where $a,b,...$ are ten-dimensional space-time indices and 
${\cal H}=d{\cal B}$ with 
${\cal B}$ the NS-NS 2-form  gauge field. The gravitational coupling 
$\kappa_{10}$ is given 
by $2\kappa^2_{10}=(2\pi)^7g_s^2\a'^4$. The ten-dimensional dilaton field 
$\phi_{10}$ 
has its zero mode substracted. 
 
For the D$p$-branes the appropriate bosonic bulk action corresponding to a 
truncation of the 
$N=2$ supergravities is 
\eqn 
\label{2.3}  
S_{bulk} = \frac{1}{2\kappa^2_{10}} \int d^{10} x \sqrt{-G} 
\left[e^{-2\phi_{10}} 
\left( R +4(\partial_a \phi_{10})^2\right)  
-\frac{1}{2\cdot (p+2)!} {\cal F}_{p+2}^2 \right]\ ,  
\eeqn  
where ${\cal F}_{p+2}=d{\cal A}_{p+1}$ and ${\cal A}_{p+1}$ is the 
$(p+1)$-form R-R gauge 
field. For $p=3$, the field equations of the action~(\ref{2.3}) must  be 
supplemented with 
the self-duality condition ${\cal F}_5=\star{\cal F}_5$, and for $p>3$ we 
consider the 
field strength ${\cal F}_{p+2}$ dual to ${\cal F}_{8-p}$.
 
We shall be interested in the overall motion of a wrapped brane in the 
uncompactified
4-dimensional space-time. In other words, we assume that the brane does not 
oscillate 
in the internal directions, and we neglect all of its worldvolume dynamics 
(except for a special case described below). 
Thus, we consider the  situation in which  the brane looks like a 
particle in the 4-dimensional space-time. It would be interesting to extend 
our analysis 
to the case of dynamical extended branes in ten uncompactified dimensions, 
including  
the world-volume excitations. We take $Y^a(\s^{\a})$ as the ten-dimensional 
target space 
coordinates and work in the static gauge 
\eqn 
\label{2.4} 
\arr{ll} 
Y^{\m}=Y^{\m}(\t)\ ,\ \ \ \ \ &\m=0,1,2,3\ ,\\ 
Y^{\hat{m}}=const.\ ,&\hat{m}=4,...,9-p\ ,\\
Y^m=\s^{m-(9-p)}\ ,\ \ \ \ \ &m=10-p,...,9\ , 
\earr 
\eeqn 
where $\s^{\a}$ are the worldvolume coordinates with $\a,\b,...=0,...,p$. We 
neglect 
all the Ka\l u\.{z}a-Klein modes in the compact and transverse directions 
$x^{\hat{m}}$. 
In the case of a compactified D$p$-brane, our discussion 
includes the possibility  of a world-line $U(1)$ gauge field  
$a_{\a}=a_{\a}(\tau )$ with an almost constant field strength $f=da$. Such a 
gauge field  
corresponds to a slowly time-varying electric field $E_{\a}=\p_{\t}a_{\a}$ 
on the
particle  world-line. The T-dual system is a D-brane moving with velocity    
$v_{\a}=E_{\a}$. Thus, we can set $a_{\a}=0$ without restricting the 
generality: a T-duality transformation on the solution describing a particle 
in nearly uniform motion  with velocity $v_{\a}$ yields the case of a slowly 
varying
electric field  $E_{\a}=v_{\a}$. The upper bound $E_{crit}$ on the magnitude
of $E_\a$ is related to the limiting velocity $c$ of the particle. The 
frame-dependent magnitude of the velocity is otherwise unrestricted.
 
With the above assumptions it is consistent to compactify the ten-dimensional 
space-time 
to 4 dimensions using the following truncation 
\eqn 
\label{2.5} 
\arr{c} 
\displaystyle{ds_{10}^2=e^{2\phi}ds_4^2+e^{2\la}dx^{\hat{m}}dx_{\hat{m}}+
e^{2\n}dx^mdx_m}\ ,\\\\ 
\displaystyle{\phi=\phi_{10}-\frac{1}{4}\ln{\hat{G}}=\phi_{10}
-\frac{1}{2}\left[ (6-p)\la+p\n\right]}\ ,\\\\ 
\displaystyle{{\cal H}=F_{\m\n}dx^{\m}\wedge dx^{\n}\wedge dx^9 
\ \ \ {\rm or}\ \ \  
{\cal F}=F_{\m\n}\e_{m_1...m_p}
dx^{\m}\wedge dx^{\n}\wedge dx^{m_1}\wedge ... \wedge dx^{m_p}}\ , 
\earr 
\eeqn 
where $\hat{G}=\det (G_{\hat{m}\hat{n}})\cdot\det (G_{mn})=
e^{2(6-p)\la+2p\n}$. 
The field $\phi$ is the dilaton field of the  
compactified theory and $ds_4$ is the line element in the Einstein frame. The 
bulk 
actions~(\ref{2.2}) and~(\ref{2.3}) reduce to 
\eqn 
\label{2.6} 
S_{bulk}=\frac{1}{2\kappa^2}\int d^4x \sqrt{-g}\left[ 
R-2\left(\p_{\m}\phi\right)^2-(6-p)\left(\p_{\m}\la\right)^2
-p\left(\p_{\m}\n\right)^2 
-\frac{1}{4}f(\phi,\la,\n)F^2\right]\ , 
\eeqn 
where $\kappa^2=\kappa^2_{10}/V_6$ defines the 4-dimensional gravitational 
coupling
with $V_6$ the volume of the internal space and 
$f(\phi,\la,\n)=e^{-a(p)\phi-b(p)\la-c(p)\n}$ with 
\eqn 
\label{2.7} 
\arr{llll} 
{\rm F-string:}\ \ \ &a(1)=2\ ,\ \ \ &b(1)=0\ ,\ \ \ &c(1)=2\ ,\\ 
{\rm D}p{\rm -brane:}\ \ \ &a(p)=0\ ,\ \ \ &b(p)=-(6-p)\ ,\ \ \ &c(p)=p\ ; 
\earr 
\eeqn 
see e.~g.~\cite{MahJHS} for a discussion of the reduction. 
 
In the case of the fundamental string ($p=1$), the action~(\ref{2.6}) admits 
the 
solution~\cite{DGHR} 
\eqn 
\label{2.8} 
\arr{c} 
ds_{4}^2=-H^{-\frac{1}{2}}dt^2+H^{\frac{1}{2}}ds^2(\bE^3)\ ,\\ 
e^{2\phi}=H^{-\frac{1}{2}}\ ,\ \ \ e^{2\la}=1\ ,\ \ \ e^{2\n}=H^{-1}\ ,\ \ \ 
F=dH^{-1}\wedge dt\ , 
\earr 
\eeqn 
where in the single centered case $H=1+\m_F/r$. The constant $\m_F$ is given 
by 
\eqn 
\label{2.9} 
\m_F=\frac{2\kappa^2}{\O_2}M=\frac{2\kappa^2}{\O_2}mT_FV_1=
m\frac{(2\pi)^6g_s^2\a'^3}{\O_2V_5}\ , 
\eeqn 
where $\O_2=4\pi$ denotes the area of the unit $2$-sphere, $m$ is the string 
winding number, 
$V_1=2\pi R_1$ is the length of the wrapped string and $V_5$ is the volume of 
the transverse
compactified space. The fundamental string tension is $T_F=(2\pi\a')^{-1}$.  
For the D$p$-branes the solution takes the form~\cite{HorStrom} 
\eqn 
\label{2.10} 
\arr{c} 
ds_{4}^2=-H^{-\frac{1}{2}}dt^2+H^{\frac{1}{2}}ds^2(\bE^3)\ ,\\ 
e^{2\phi}=1\ ,\ \ \ e^{2\la}=H^{\frac{1}{2}}\ ,\ \ \ e^{2\n}=
H^{-\frac{1}{2}}\ ,\ \ \ 
F=dH^{-1}\wedge dt\ , 
\earr 
\eeqn 
where $H=1+\m_p/r$. The constant  $\m_p$ is given by 
\eqn 
\label{2.11}
\m_p=\frac{2\kappa^2}{\O_2}M=\frac{2\kappa^2}{\O_2}NT_pV_p=
N\frac{(2\pi)^{7-p}g_s\a'^{\frac{7-p}{2}}}{\O_2V_{6-p}}\ ,
\eeqn 
where $N$ is the number of D-branes or the D-brane winding number, and the 
D$p$-brane  
tension is $T_p=\left((2\pi)^p\a'^{\frac{p+1}{2}}g_s\right)^{-1}$.
 
Next we consider the action for the brane source. In the case of the 
fundamental string, 
we start with the bosonic action 
\eqn 
\label{2.12} 
S_F=-\frac{T_F}{2}\int d^2\s 
\left( \sqrt{-\g}\g^{\a\b}\p_{\a}Y^a\p_{\b}Y^bG_{ab} 
+\epsilon^{\a\b}\p_{\a}Y^a\p_{\b}Y^b{\cal B}_{ab}\right)\ . 
\eeqn 
With the Ans\"{a}tze~(\ref{2.4}) and~(\ref{2.5}) this action yields the 
following particle action: 
\eqn 
\label{2.13} 
S=-M\int d\t\left[\frac{1}{2} 
\left(-e^{2\phi}U^{\m} U^{\n}g_{\m \n} +e^{2\n}\right)+U^{\m}C_{\m}\right]\ , 
\eeqn 
where $dC=F$ and we have defined the transverse 4-velocity vector  
\begin{equation} 
 U^\m (\t ) \equiv \frac{d Y^\mu }{d\tau} .  
\end{equation} 
In the case of a D$p$-brane, we start with the effective action 
\eqn 
\label{2.14} 
\arr{rcl} 
S_{Dp}&=&\displaystyle{-T_p\int d^{p+1}\s 
\left( e^{-\phi_{10}}\sqrt{-\det\left(\hat{G}_{\a\b}+2\pi\a'f_{\a\b}-
\hat{\cal B}_{\a\b}\right)} 
\phantom{\frac{1}{(p+1)!}}\right.}\\ 
&&\displaystyle{\ \ \ \ \ \ \ \ \ \ \ \ \ \ \ \ \ \ \ \ \ \ \  
\left.+\frac{1}{(p+1)!}\epsilon^{\a_1...\a_{p+1}}
\hat{\cal A}_{\a_1...\a_{p+1}}\right)}\ , 
\earr 
\eeqn 
where $\hat{G}_{\a\b}$, $\hat{\cal B}_{\a\b}$ and 
$\hat{\cal A}_{\a_1...\a_{p+1}}$ are 
the pull-backs to the worldvolume of the ten-dimensional bulk fields 
$G_{ab}$,  
${\cal B}_{ab}$ and ${\cal A}_{a_1...a_{p+1}}$, respectively. Note that the 
expansion of  
the Dirac-Born-Infeld part of the action around flat space is of second order 
in 
${\cal B}_{\a\b}$. Thus, in the linearised approximation considered in this 
paper we can  
set this field to zero. Also, as explained above, we can set $f_{\a \b}=0$. 
The Dirac-Born-Infeld part of the action becomes 
\eqn 
\label{2.15} 
\arr{rcl} 
S&=&\displaystyle{-T_p\int d^{p+1}\s~e^{-\phi_{10}}
\sqrt{-det\left(\hat{G}_{\a\b}\right)}}\\ 
&=&  
\displaystyle{-\frac{T_p}{2}\int d^{p+1}\s~e^{-\phi_{10}}\sqrt{-\g} 
\left( \g^{\a\b}\p_{\a}Y^a\p_{\b}Y^bG_{ab}-(p-1)\right)}\ . 
\earr 
\eeqn 
Note that,  in the effective action~(\ref{2.14}), we have not included the 
generalized 
Wess-Zumino coupling of the form 
${\cal A} \exp {\cal F}$~\cite{MRD,GHT}. This is 
consistent with our assumptions. For example, in the case of a time-varying 
electric 
field we should consider terms of the form $f\wedge {\cal A}_{p-1}$, and 
the D$p$-brane 
would emit ${\cal A}_{p-1}$ gauge field radiation. Also, we have neglected 
certain 
anomalous couplings~\cite{GHM} and curvature terms~\cite{BBG} which will 
play no role 
in the analysis of the present paper. With the Ans\"{a}tze~(\ref{2.4}) 
and ~(\ref{2.5}) 
we obtain the following particle action for the wrapped D$p$-brane 
\eqn 
\label{2.16} 
S=-M\int d\t\left[\frac{1}{2}~e^{-\frac{6-p}{2}\la+\phi-\frac{p}{2}\n} 
\left(-U^\m U^\n g_{\m \n} +p\ e^{-2\phi+2\n}-(p-1)e^{-2\phi}\right) 
+U^{\m}C_{\m}\right]\ . 
\eeqn 
 
\subsection{Field Equations} 
 
Here we give the bosonic field equations for the coupled supergravity-brane 
system. 
Consider first the field equations for the scalars. In the case of the 
F-string, 
the field equation for the dilaton field is
\eqn 
\label{2.18} 
\Box\phi+\frac{1}{8}f(\phi,\n)F^2=\frac{1}{2}\kappa^2M \int d\t~e^{2\phi}~ 
\frac{\d^{(4)}(x-Y(\t))}{\sqrt{-g}}\ , 
\eeqn 
while in the case of the D$p$-brane it  takes the form 
\eqn 
\label{2.19}
\Box \phi=\frac{1}{4}\kappa^2M \int d\t~e^{-\frac{6-p}{2}\la+\phi-
\frac{p}{2}\n} 
\left( 1-p~e^{-2\phi+2\n}+(p-1)e^{-2\phi}\right) 
\frac{\d^{(4)}(x-Y(\t))}{\sqrt{-g}}\ . 
\eeqn 
The  field equation for the scalar $\n$ in the case of the F-string is 
\eqn 
\label{2.20} 
\Box\n +\frac{1}{4}f(\phi,\n)F^2=
\kappa^2M \int d\t~e^{2\n}~\frac{\d^{(4)}(x-Y(\t))}{\sqrt{-g}}\ , 
\eeqn 
while in the case of the D$p$-brane it is 
\eqn 
\label{2.21} 
\arr{c} 
\displaystyle{\Box\n +\frac{1}{8}f(\la,\n)F^2=}\\ 
\displaystyle{\frac{1}{4}\kappa^2M 
\int d\t~e^{-\frac{6-p}{2}\la+\phi-\frac{p}{2}\n}
\left(-1+(4-p)e^{-2\phi+2\n} 
+(p-1)e^{-2\phi}\right) 
\frac{\d^{(4)}(x-Y(\t))}{\sqrt{-g}}}\ . 
\earr 
\eeqn 
In the case of the F-string, the field equation for  the scalar field $\la$ 
simplifies to 
\eqn
\label{2.22} 
\Box\la=0\ ,
\eeqn
while in the case of the D$p$-brane it reads
\eqn 
\label{2.24} 
\arr{c} 
\displaystyle{\Box\la -\frac{1}{8}f(\la,\n)F^2=}\\ 
\displaystyle{-\frac{1}{4}\kappa^2M 
\int d\t~e^{-\frac{6-p}{2}\la+\phi-\frac{p}{2}\n}
\left(1+p~e^{-2\phi+2\n} 
-(p-1)e^{-2\phi}\right) 
\frac{\d^{(4)}(x-Y(\t))}{\sqrt{-g}}}\ . 
\earr 
\eeqn

The gauge field equation for both the F-string and  the D$p$-brane is 
\eqn 
\label{2.25} 
\p_{\m}\left( \sqrt{-g}f(\phi,\la,\n)F^{\m\n}\right)= 
2\kappa^2 M\int d\t U^{\n}(\t)\d^{(4)}(x-Y(\t))\ . 
\eeqn 

Finally, the Einstein equations read 
\eqn 
\label{2.26} 
R^{\m\n}-\frac{1}{2}g^{\m\n}R=\kappa^2 
\left( T^{\m\n}_{\phi}+T^{\m\n}_{\n}+T^{\m\n}_{C}+
T^{\m\n}_{\la}+T^{\m\n}_{source}\right)\ , 
\eeqn 
where
\eqn 
\label{2.27} 
\arr{l} 
\displaystyle{\kappa^2T^{\m\n}_{\phi}= 
2\left(\p^{\m}\phi\p^{\n}\phi-\frac{1}{2}g^{\m\n}(\p \phi)^2\right)}\ ,\\ 
\displaystyle{\kappa^2T^{\m\n}_{\n}= 
p\left(\p^{\m}\n\p^{\n}\n-\frac{1}{2}g^{\m\n}(\p \n)^2\right)}\ ,\\ 
\displaystyle{\kappa^2T^{\m\n}_{\la}=
(6-p)\left(\p^{\m}\la\p^{\n}\la-\frac{1}{2}g^{\m\n}(\p \la)^2\right)}\ ,\\ 
\displaystyle{\kappa^2T^{\m\n}_{C}=\frac{1}{2} f(\phi,\la,\n) 
\left(F^{\m}_{\ \eta} F^{\n\eta}-\frac{1}{4}g^{\m\n}F^2\right)}\ , 
\earr 
\eeqn 
are the contributions of the scalar fields and of the gauge field to the 
stress-energy 
tensor derived from~(\ref{2.6}).  
The stress-energy tensors of the sources are given by 
\eqn 
\label{2.28} 
\arr{ll} 
{\rm F-string:}\ \ \ & 
\displaystyle{T^{\m\n}_{source}=
M\int 
d\t U^{\m}(\t)U^{\n}(\t)~e^{2\phi}~\frac{\d^{(4)}(x-Y(\t))}{\sqrt{-g}}}\ ,\\ 
{\rm D}p{\rm -brane:}\ \ \ & 
\displaystyle{T^{\m\n}_{source}=
M\int d\t U^{\m}(\t)U^{\n}(\t)~e^{-\frac{6-p}{2}\la+\phi-\frac{p}{2}\n}~
\frac{\d^{(4)}(x-Y(\t))}{\sqrt{-g}}}\ . 
\earr 
\eeqn 
 
\subsection{Accelerated Branes}

For small string coupling (and $r$ large enough), the supergravity 
solutions~(\ref{2.8})
and~(\ref{2.10}) can be written as an expansion in $(\kappa^2M)$ around flat 
space. Then,
we may regard the branes as  delta-function sources localised at  $r=0$. In 
order to accelerate the 
branes we further consider some weak external field. For simplicity we 
consider an 
external scalar field and to be definite we restrict ourselves to the scalar 
field
$\n$. If $\n_0(x)$ is the external accelerating field we write
\eqn
\label{2.29} 
\n(x)\rightarrow \n_0(x)+\n(x)\ ,
\eeqn
where $\n_0(x)$ admits some Fourier decomposition 
$\n_0(x)=\int d^4k \n_0(k) e^{-ik\cdot x}$ and  the field
$\n(x)$ represents the scalar field generated by the brane source.

To leading order in the external field $\n_0(x)$ the equation of motion for 
the 
F-string is
\eqn
\label{2.30} 
\dot{U}^\m+\p^\m\n_0=0\ ,
\eeqn
where $\dot{U}\equiv dU/d\t$ is the 4-acceleration. For the D$p$-brane, the 
corresponding equation is
\eqn
\label{2.31} 
\dot{U}^\m+\frac{p}{2}\left( -U^\m U\cdot\p\n_0+\p^\m\n_0\right)=0\ .
\eeqn
In both cases, taking the scalar product of the 4-velocity with the 
corresponding 
equation of motion, we conclude that $U\cdot\p\n_0=0$. Thus, (\ref{2.30})
simplifies to
\eqn
\label{2.32} 
\dot{U}^\m+\frac{p}{2}\p^\m\n_0=0\ .
\eeqn
Also, note that the acceleration is of first order in $\n_0$, i.e. 
$\dot{U}\sim\n_0$.
In the non-relativistic limit the equations of motion for the F-string and 
D$p$-brane become
\eqn
\label{2.33}
\vdot+\p\n_0=0\ \ \ {\rm and}\ \ \ \vdot+\frac{p}{2}\p\n_0=0\ ,
\eeqn
respectively.

\subsection{Linearised Field Equations}

Since the fields generated by the brane source can be written as an expansion 
in $(\kappa^2M)$, we shall work to first order in this expansion parameter and 
keep only the terms linear in such fields.  In other words, the source terms 
in the field equations of section 2.2 are taken to be small. Also, we assume 
that the  external field 
$\n_0(x)$ is  weak and keep only the  terms linear in this accelerating field.
Since the magnitude $a$ of the acceleration is limited on dimensional grounds
to $a\ll 1/\sqrt{\alpha '}$ in order for the higher in $\alpha '$  derivative 
corrections to supergravity to be negligible, the external field $\n_0 (x)$ 
must satisfy $(\p \n_0 )^2 \ll 1/\a'$. In the case of D-branes,  the upper 
bound $1/\sqrt{\alpha '}$ on the magnitude of the acceleration also ensures
that corrections due to derivatives of the world-volume fields 
(in particular derivatives of 
the scalars $Y$) to the 
effective brane
action can be consistently neglected.   

In the case of the  F-string, the linearised field equations  
for the scalar fields  take the form 
\eqn
\label{2.34}
\arr{c}
\displaystyle{\p^2\phi=\frac{1}{2}\kappa^2M\int d\t~\d^{(4)}(x-Y(\t))}\ ,\\\\
\displaystyle{\p^2\n=\kappa^2M\int d\t 
\left( 1+2\n_0^{\phantom{1}}(Y(\t))\right)
\d^{(4)}(x-Y(\t))}\ ,\\\\
\p^2\la=0\ .
\earr
\eeqn
In the case of the D$p$-brane we find
\eqn
\label{2.35}
\arr{c}
\displaystyle{\p^2\phi=\frac{1}{2}\kappa^2M\int d\t 
\left(-p\ \n_0^{\phantom{1}}(Y(\t))\right)\d^{(4)}(x-Y(\t))}\ ,\\\\
\displaystyle{\p^2\n=\frac{1}{2}\kappa^2M\int d\t 
\left(1+\left(4-\frac{3}{2}p\right)\n_0(Y(\t))\right)\d^{(4)}(x-Y(\t))}\ ,\\\\
\displaystyle{\p^2\la=-\frac{1}{2}\kappa^2M\int d\t 
\left(1+\frac{p}{2}\n_0(Y(\t))\right)\d^{(4)}(x-Y(\t))}\ .
\earr
\eeqn

The linearised gauge field equation for both the F-string and the D$p$-brane 
is given 
by
\eqn
\label{2.36}
\p_\m\left( \left(1-c(p)\n_0^{\phantom{1}}(x)\right)F^{\m\n}\right)=
2\kappa^2M\int d\t U^\n(\t)\d^{(4)}(x-Y(\t))\ .
\eeqn

Finally, we give the field equations for the metric fluctuations. 
In the case of the F-string, we find
\eqn
\label{2.37}
\arr{rcl}
\p^2\bar{h}^{\m\n}&=&
\displaystyle{-2\kappa^2M\int d\t U^\m(\t)U^\n(\t)\d^{(4)}(x-Y(\t))}\\\\
&&
\displaystyle{-2p\left(2\p^{(\m}\n_0\p^{\n)}\n-\eta^{\m\n}\p\n_0
\cdot\p\n\right)}\ ,
\earr
\eeqn
where the metric was split as $g_{\m\n}=\eta_{\m\n}+h_{\m\n}$ and the 
definition 
$\bar{h}_{\m \n}\equiv h_{\m \n}-\frac{1}{2}\eta_{\m \n} h$ with 
$h\equiv h^{\m}_{\ \m}$ was introduced. The tensor $\bar{h}_{\m\n}$ is taken 
to satisfy  the gauge condition  
\eqn
\p_{\m}\bar{h}^{\m\n}=0. \label{gaugecond}
\eeqn
 In the  case of the D$p$-brane,  the corresponding equation is
\eqn
\label{2.38}
\arr{rcl}
\p^2\bar{h}^{\m\n}&=&
\displaystyle{-2\kappa^2M\int d\t U^\m(\t)U^\n(\t)
\left(1-\frac{p}{2}\n_0(Y(\t))\right)\d^{(4)}(x-Y(\t))}\\\\
&&
\displaystyle{-2p\left(2\p^{(\m}\n_0\p^{\n)}\n-
\eta^{\m\n}\p\n_0\cdot\p\n\right)}\ ,
\earr
\eeqn

To understand physically the various terms in the above equations, consider 
equation~(\ref{2.38}). The first line on the right hand side has two parts:  
the factor of 1 inside the 
bracket gives  a source term for the brane, while the factor involving
$\n_0(Y(\t))$ governs the scattering of the external field by the brane source,
i.e. the  absorption of closed strings with the quantum numbers of the 
field $\n_0(x)$ and the emission of
closed strings with those of the  graviton. The second line on the right hand 
side of~(\ref{2.38})
governs the interaction of the external field $\n_0(x)$ with the field $\n$ 
generated by the brane 
source, which results in the emission of gravitons.
 
\section{Scalar Radiation from Accelerated Branes}  
\news 

\label{scalarrad}
 
In this section we study the scalar field radiation emitted by accelerated 
branes. 
There are three
different scalar fields, namely the dilaton $\phi$ and the compactification 
scalars
$\la$ and $\n$. For the sake of brevity, we will only  consider the scalar 
field $\n$;
similar results hold for the other scalars. The results obtained below  are 
generalizations of standard results in classical electrodynamics; for a more 
detailed 
discussion, see e.g.~\cite{JDJ}, in particular chapter 14. We start by 
finding a 
solution to the field equation for the scalar field $\n$. Next we compute the 
energy 
flux of the emitted radiation and study its angular dependence. This result 
is used 
to calculate the Thompson scattering cross sections for the scalar $\n$, from 
which 
the  classical radii of the branes are deduced.

The retarded solution to the linearised field equations~(\ref{2.34}) 
and~(\ref{2.35}) 
for the scalar $\n(x)$ can be written in the integral form
\eqn
\label{3.1} 
\n(x)=\int d^D x' D_r(x-x')J(x')\ , 
\eeqn 
where $D_r(x-x')$ is the retarded Green function, 
\eqn 
\label{3.2} 
D_r(x-x')=-\frac{1}{2\pi}\th\left(x^0-x'^0 \right)\d\left[(x-x')^2\right] 
\eeqn 
(here $\theta (x_0 )$ is the usual step function satisfying 
$\theta (x_0 ) =0$ for 
$x_0 <0$ and $\theta (x_0 ) =1$ for $x_0 >0$), and 
\eqn 
\label{3.3} 
J(x')=A\kappa^2 M \int d\t 
\left(1+d(p)\n_0^{\phantom{1}}(Y(\t))\right)\d^{(4)}(x'-Y(\t))\ , 
\eeqn 
with 
\eqn 
\label{3.4} 
\arr{lll} 
{\rm F-string:}\ \ \ & A=1\ ,\ \ \ &d(1)=2\ ,\\
{\rm D}p{\rm -brane:}\ \ \ &\displaystyle{A=\frac{1}{2}}\ ,\ \ \ &
\displaystyle{d(p)=4-\frac{3}{2}p}\ .\\
\earr 
\eeqn 
The retarded Green function~(\ref{3.2}) is non-zero only on the forward 
light-cone of the source
point, a requirement that leads to the solution 
\eqn
\label{3.5}
\n(x)=\frac{A\kappa^2 M}{4\pi}
\left[\frac{1+d(p)\n_0(Y(\t))}{U\cdot(x-Y(\t))}\right]_{\t=\t_0}\ ,
\eeqn
where $\t_0$ is defined by the light cone condition 
\eqn 
\label{3.6} 
\left[x-Y(\t_0)\right]^2=0\ ,
\eeqn 
and the retardation requirement $x^0>Y^0(\t_0)$. The light cone condition is 
equivalent to  
\eqn 
\label{3.7} 
x^0-Y^0(\t_0)=|\bx-\bY(\t_0)|\equiv R\ . 
\eeqn 

\subsection{Radiated Power}

To calculate the power radiated by the brane in the form of waves of the 
scalar field $\n$, 
we define the Poynting vector $\bS$ by $S^i=T^{0i}_{\n}$, where $T^{\m\n}_{\n}$
is the contribution of this scalar field to the stress-energy tensor given 
in~(\ref{2.27}).
In the linear approximation,
\eqn
\label{3.8} 
S^i_\n=T_\n^{0i}=\frac{p}{\kappa^2}\left(\p^0\n\right)\left(\p^i\n\right)\ .
\eeqn 
Notice that we assume the incident beam $\n_0(x)$ to be well localised such 
that
it does not mix with the emitted radiation far from the source (otherwise 
there would
be  terms proportional to $(\p\n)(\p\n_0)$ in the Poynting vector).

To calculate the Poynting vector we have to compute the derivatives with 
respect 
to the observation coordinates $x\equiv(t,\bx)$ of the retarded 
quantity~(\ref{3.5}). 
In doing so it is important to realize that functions of the proper time 
$\t$ depend 
implicitly on $x$ through the relation~(\ref{3.6})~\cite{LL}. For a general 
function 
$f(x,\t)$ we have 
\eqn 
\label{3.9} 
\p^{\m}f=\p^{\m}f\Big|_{\t}+\p^{\m}\t\frac{df}{d\t}\ , 
\eeqn 
where 
\eqn 
\label{3.10} 
\p^{\m}\t=\frac{(x-Y(\t))^{\m}}{U\cdot(x-Y(\t))}\ . 
\eeqn 
Thus, the derivative of the scalar field $\n$ is 
\eqn
\label{3.11} 
\arr{rcl}
\p^\m\n&=&\displaystyle{\frac{A\kappa^2M}{4\pi}\frac{1}{[U\cdot(x-Y(\t))]^2}
\left\{\left[1+d(p)\n_0^{\phantom{1}}(Y(\t))\right]
\left(-\frac{(x-Y(\t))^\m}{U\cdot(x-Y(\t))}\right.\right.}\\\\
&&\displaystyle{\left.\left.
-U^\m-\frac{(x-Y(\t))^\m}{U\cdot(x-Y(\t))}\dot{U}\cdot(x-Y(\t))\right)
+(x-Y(\t))^\m d(p)\p\n_0\cdot U\right\}}\ ,
\earr
\eeqn
evaluated at the retarded time $\t=\t_0$. The last term in this equation 
vanishes 
because $\p\n_0\cdot U=0$. It is convenient to write
$\p^\m\n$ in non-covariant form using the formulae 
\eqn 
\label{3.12} 
\arr{c} 
\displaystyle{(x-Y)^{\m}\equiv R (1,\bn)\ ,\ \ \  
U^{\m}= \g(1,\bv)\ ,\ \ \  
\g^2=\displaystyle{\frac{1}{1-\bv^2}}}\ ,\\ 
\dot{U}^{\m}\equiv \displaystyle{\frac{dU^{\m}}{d\t}} = 
\g^2\left(\g^2\bv\cdot\vdot,\vdot+\g^2\bv(\bv\cdot\vdot)\right)\ , 
\earr 
\eeqn 
where $\bn$ is the unit vector in the  direction of observation 
$\bx-\bY(\t)$ and  
$\vdot=d\bv/dY^0$. The result is
\eqn
\label{3.13} 
\arr{rcl}
\p\n& = &
\displaystyle{\frac{A\kappa^2M}{4\pi}\left(1+d(p)\n_0^{\phantom{1}}(Y(\t))
\right)
\left[\frac{(1,\bn)}{R^2\g^3(1-\bn\cdot\bv)^3}-\frac{(1,\bv)}{R^2\g(1-\bn
\cdot\bv)^2}
\right.}\\\\
&&\displaystyle{\left.\ \ \ \ \ \ \ \ \ \ \ \ \ \ \ \ \ \ \ \ \ \ \ \ \ \ \ \ 
-\frac{(1,\bn)}{R\g(1-\bn\cdot\bv)^2}\left(\g^2(\bv\cdot\vdot)
-\frac{\bn\cdot\vdot}{1-\bn\cdot\bv}\right)\right]}\ .
\earr
\eeqn
This divides naturally into `velocity fields' and `acceleration fields'. The
first two terms in the square bracket are  velocity fields which fall off as
$1/R^2$ and are static fields, as can be seen by 
performing a Lorentz transformation. The last term in the bracket contains 
the acceleration fields, which fall off as $1/R$ and are typical radiation 
fields.
Both the velocity and the acceleration fields are multiplied by an extra 
factor of
$d(p)\n_0(Y(\t))$ which is induced by the external field. Keeping only the 
radiation
fields, and noting that $\dot{U}\sim\n_0$, we find to linear order
\eqn
\label{3.14} 
\p\n = \frac{A\kappa^2M}{4\pi}
\left[-\frac{(1,\bn)}{R\g(1-\bn\cdot\bv)^2}\left(\g^2(\bv\cdot\vdot)
-\frac{\bn\cdot\vdot}{1-\bn\cdot\bv}\right) \right]\ .
\eeqn
Thus, $\p^i\n=(\p^0\n)n^i$ for the radiation fields and hence
\eqn 
\label{3.15} 
\bS=\frac{p}{\kappa^2}\left(\p^0\n\right)^2\bn\ . 
\eeqn 
For an accelerated brane in a frame where the velocity is 
small ($v \ll 1$), we find
\eqn 
\label{3.16} 
\p^0\n=\frac{A\kappa^2 M}{4\pi}\frac{\bn\cdot\vdot}{R}\ , 
\eeqn 
which gives the Poynting vector 
\eqn 
\label{3.17} 
\bS=p\left(\frac{A\kappa M}{4\pi R}\right)^2 
|\vdot|^2\cos^2{\th}\ \bn\ , 
\eeqn 
where $\th$ is the angle between $\bv$ and the observation direction $\bn$.
The power radiated per unit of solid angle is  
$\frac{dP(t)}{d\O}=R^2(\bS\cdot\bn)$. Upon integrating over all  
solid angle we obtain the total (instantaneous) power radiated,
\eqn 
\label{3.18} 
P=\frac{p}{12\pi}(A\kappa M)^2|\vdot|^2\ . 
\eeqn 
This is the Larmor formula for a (nonrelativistic) accelerated wrapped brane. 
The total power emitted by an accelerated wrapped brane in relativistic 
motion is 
given by the unique Lorentz invariant generalization of~(\ref{3.18}) obtained 
by 
substituting 
\eqn 
\label{3.19} 
|\vdot|^2\rightarrow \dot{U}\cdot\dot{U}= 
\g^4\left(\g^2(\bv\cdot\vdot)^2+\vdot^2\right)\ . 
\eeqn 

\subsection{Angular Distribution of Radiation} 
 
\begin{figure}
\begin{center}
\includegraphics{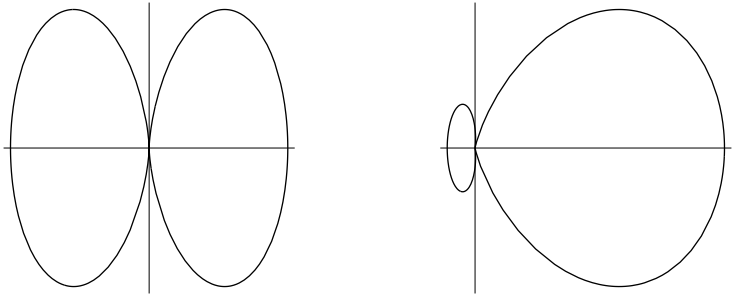}
\end{center}
\caption{Angular distribution of the scalar field radiation for an 
accelerated brane 
with $\bv$ and $\vdot$ {\em parallel} in the non-relativistic (left) and the 
relativistic (right) regimes. Note that as $v \to 1$, the angular 
distribution is 
tipped forward and increases in magnitude.} \label{figu1}
\end{figure}

In terms of the  proper time $t'$ of the brane, the general formula for the 
power 
radiated in the form of waves of  the scalar field $\n$  per unit of 
solid angle is given  by  
\eqn 
\label{3.20} 
\frac{dP(t')}{d\O}=p\left(\frac{A\kappa M}{4\pi}\right)^2 
\frac{\left[\g^2(1-\bn\cdot\bv)\bv\cdot\vdot-\bn\cdot\vdot\right]^2} 
{(1-\bn\cdot\bv)^5}\ . 
\eeqn 
We consider two special cases of this result: in the first, $\bv$ and $\vdot$ 
are parallel,
while in the second they are  perpendicular. 
In the former situation,  equation~(\ref{3.20}) becomes 
\eqn 
\label{3.21} 
\frac{dP(t')}{d\O}=p\left(\frac{A\kappa M}{4\pi}\right)^2 
\g^2\dot{v}^2\frac{(v-\cos{\th})^2}{(1-v\cos{\th})^5}\ . 
\eeqn 
For $v\ll 1$, this is the Larmor result. For
$v\sim 1$, the angular distribution is tipped 
forward as indicated schematically in  figure~\ref{figu1}. 

\begin{figure}
\begin{center}
\includegraphics{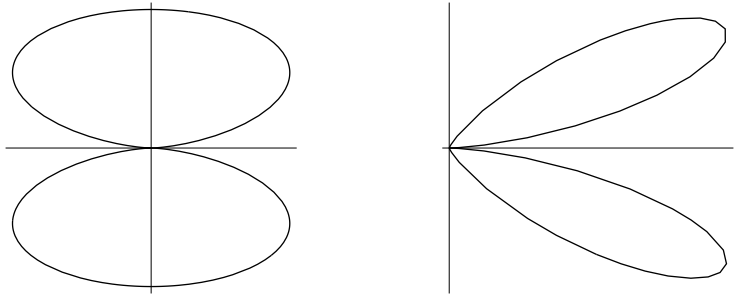}
\end{center}
\caption{Angular distribution of the scalar field radiation for an 
accelerated brane with 
$\bv$ and $\vdot$ {\em perpendicular} at fixed angle $\phi$ (with $\cos \phi 
\neq 0$) in 
the non-relativistic (left) and the relativistic (right) regimes. As 
$v \to 1$, the 
distribution is  tipped forward and the intensity reaches a maximum for a 
particular value 
of the angle, $\theta_{max}$. In the direction perpendicular to $\vdot$  
($\cos \phi =0$), 
no radiation is observed.} \label{figu2}
\end{figure}

If $\bv$ and $\vdot$ are perpendicular, then  the angular 
distribution~(\ref{3.20}) becomes 
\eqn 
\label{3.22} 
\frac{dP(t')}{d\O}=p\left(\frac{A\kappa M}{4\pi}\right)^2 
\frac{\dot{v}^2}{\g^2}\frac{\sin^2{\th}\cos^2{\phi}}{(1-v\cos{\th})^5}\ , 
\eeqn 
where we are using spherical coordinates such that $\bv\cdot\bn=v\cos{\th}$ and
$\vdot\cdot\bn=\dot{v}\sin{\th}\cos{\phi}$. The    distribution in $\theta$  
at fixed angle $\phi$ is shown schematically in figure~\ref{figu2}.

\subsection{Thompson Scattering}

The previous results can be used to calculate the cross section for the 
emission
of waves of the scalar field $\n$ (as well as radiation of $\phi$ and $\la$ 
waves). In the 
non-relativistic
limit the brane motion has the same frequency as the incident radiation and the
process can be described as scattering of the incident radiation. For 
non-relativistic motion, the instantaneous  power per unit of solid angle 
radiated by the brane is 
\eqn
\label{3.23} 
\frac{dP}{d\O}=p\left(\frac{A\kappa M}{4\pi}\right)^2|\vdot|^2\cos^2{\th}\ .
\eeqn
The incident wave $\n_0(k)e^{-ik\cdot x}$ induces an acceleration
\eqn
\label{3.24} 
\vdot=-pAi\bk\n_0(k)e^{-ik\cdot x}\ .
\eeqn
In the long wave-length approximation, we can assume that the brane moves a 
negligible
part of a wave-length during one oscillation and therefore the time average 
of $|\vdot|^2$
is $\frac{1}{2}Re(\vdot\cdot\vdot^*)$. Then, the average radiated power per 
unit of solid 
angle is
\eqn
\label{3.25} 
\left\langle\frac{dP}{d\O}\right\rangle=p\left(\frac{A\kappa M}{4\pi}\right)^2
\frac{(pA)^2}{2}|\bk|^2|\n_0|^2\cos^2{\th}\ .
\eeqn
Since the average incident flux for the plane wave is 
$\frac{p}{\kappa^2}\frac{1}{2}|\bk|^2|\n_0|^2$
we obtain the differential scattering cross section 
\eqn
\label{3.26} 
\frac{d\s}{d\O}=
\left(\frac{pA^2\kappa^2 M}{4\pi}\right)^2\cos^2{\th}\ .
\eeqn
This result is in agreement with expectations from consideration of  tree 
level string diagrams.
For the F-string, the scattering amplitude is proportional to 
$(\kappa^2M)\sim g_s^2$, corresponding to a sphere with four vertex operator 
insertions.
In the case of the D$p$-brane, the amplitude is proportional to 
$(\kappa^2M)\sim g_s$, 
corresponding to a disk with two vertex operator insertions in  the interior.

Integrating the differential cross section~(\ref{3.26}) over all solid angle, 
we obtain 
the Thompson cross section for  emission of waves of the scalar 
field $\n$. The length scale 
$\kappa^2 M$ defines the  classical radii $R_F$ and $R_p$ of the F-string 
and the D$p$-branes:
\eqn
\label{3.27} 
R_F\sim \kappa^2 M_F\sim g^2_s\frac{\a'^3}{V_5}\ \ \ {\rm and}\ \ \ 
R_p\sim \kappa^2 M_p\sim g_s\frac{\a'^{\frac{7-p}{2}}}{V_{6-p}}\ .
\eeqn

\section{Gauge Field Radiation from Accelerated Branes}
\news

Consider the linearised field equation for the gauge field,
\eqn
\label{4.1}
\p_\m\left( \left(1-c(p)\n_0^{\phantom{1}}(x)\right)F^{\m\n}\right)
=2\kappa^2M\int d\t U^\n(\t)\d^{(4)}(x-Y(\t))\ .
\eeqn
In order to solve this coupled equation, we split the gauge field 
as $C^\m=C^\m_{(0)}+C^\m_{(1)}$, where $C^\m_{(0)}$ does not depend explicitly 
on $\n_0$ and $C^\m_{(1)}$ depends linearly on $\n_0$. Working in the gauge
\begin{equation}
\p_\m C^\m=0 \label{Cgauge}
\end{equation}
and dropping quadratic terms in $\n_0$, we find the pair of equations
\eqn
\label{4.2}
\arr{c}
\displaystyle{\p^2C^\m_{(0)}=2\kappa^2 M
\int d\t U^\n(\t)\d^{(4)}(x-Y(\t))}\ , \\
\displaystyle{\p^2C^\m_{(1)}=c(p)\p_\m\left(\n_0(x)F^{\m\n}_{(0)}\right)}\ .
\earr
\eeqn
As for the scalar fields, we use the retarded Green function~(\ref{3.2}) to 
find the solution
for $C^\m_{(0)}$. The result is 
\eqn
\label{4.3}
C^\m_{(0)}=\frac{\kappa^2 M}{2\pi}
\left[\frac{U^\n(\t)}{U\cdot(x-Y(\t))}\right]_{\t=\t_0}\ ,
\eeqn
and the corresponding field strength is
\eqn
\label{4.4}
F^{\m\n}_{(0)}=\frac{\kappa^2M}{2\pi}
\frac{2(x-Y(\t))^{[\m}}{[U\cdot(x-Y(\t))]^2}
\left[-\frac{U^{\n]}}{U\cdot(x-Y(\t))}
-\frac{U^{\n]}\dot{U}\cdot(x-Y(\t))}{U\cdot(x-Y(\t))}+\dot{U}^{\n]}\right]\ ;
\eeqn
this expression must be evaluated at $\t=\t_0$. This is  the same answer as in 
classical electrodynamics. Since $C^\m_{(0)}$ obeys by itself the gauge 
condition 
$\p_\m C^\m_{(0)}=0$, the field equation for $C^\m_{(1)}$ becomes
\eqn
\label{4.5}
\p^2C^\m_{(1)}=c(p)\left( (\p_\m\n_0)F^{\m\n}_{(0)}+
\n_0(x)\p^2C^\n_{(0)}\right)\ ,
\eeqn
which can be written as
\eqn
\label{4.6}
\p^2C^\m_{(1)}=c(p)\left( (\p_\m\n_0)F^{\m\n}_{(0)}
+\n_0(x) 2\kappa^2 M\int d\t U^\n(\t)\d^{(4)}(x-Y(\t))\right)\ .
\eeqn
Since $\dot{U}\sim\n_0$, we may drop the acceleration terms in 
$F^{\m\n}_{(0)}$. 
Then, to first order in $\n_0$ equation (\ref{4.6}) simplifies to
\eqn
\label{4.7}
\arr{rcl}
\p^2C^\m_{(1)}&=&\displaystyle{\frac{\kappa^2M}{2\pi}c(p)
\left[-(\p_\m\n_0)\left(\frac{2(x-Y(\t))^{[\m}U^{\n]}}{[U\cdot (x-Y(\t))]^3}
\right)_{\t=\t_0}
\right.}\\\\
&&\displaystyle{\left.\ \ \ \ \ \ \ \ \ \ \ \ \ \ \ 
+4\pi\int d\t~\n_0(Y(\t))U^\n(\t)\d^{(4)}(x-Y(\t))\right]}\ .
\earr
\eeqn
Let us first consider the second term on the right hand side of this equation.
The corresponding contribution to the field $C^\m_{(1)}$ is
\eqn
\label{4.8}
C^\m_{(1)}\sim\frac{\kappa^2 M}{2\pi}c(p)
\left[\frac{\n_0(Y(\t))U^\n(\t)}{U\cdot (x-Y(\t))}\right]_{\t=\t_0}\ ,
\eeqn
which gives a field strength
\eqn
\label{4.9}
F^{\m\n}_{(1)}\sim\frac{\kappa^2 M}{2\pi}c(p)
\left[\frac{2(\p\n_0\cdot U)(x-Y(\t))^{[\m}U^{\n]}}{[U\cdot (x-Y(\t))]^2}
\right]_{\t=\t_0}+ \ldots
\eeqn
where the ellipsis refers to  velocity terms that are linear in $\n_0$ but 
fall off
as $1/R^2$ and to  acceleration terms that fall off as $1/R$ but are quadratic
in $\n_0$. Since $\p\n_0\cdot U=0$ we conclude that the contribution to the 
observed radiation of the second term on the right hand side of~(\ref{4.7}) 
is negligible. 
Thus, within our approximations equation (\ref{4.7}) reduces to
\eqn
\label{4.10}
\p^2C^\n_{(1)}=-\frac{\kappa^2 M}{2\pi}c(p)(\p_\m\n_0)
\left(\frac{2(x-Y(\t))^{[\m}U^{\n]}}{[U\cdot (x-Y(\t))]^3}\right)_{\t=\t_0}\ .
\eeqn
Physically this equation describes the interaction of the external field 
$\n_0(x)$
with the velocity fields in $C^\m_{(0)}$. This generates  the field 
$C^\m_{(1)}$. 
The contribution of $C^\m_{(1)}$ to the  energy flux at infinity is not 
negligible. 
From equation~(\ref{4.10}), we conclude that $\p^2C^\n_{(1)}\sim\n_0/R^2$
and therefore $\p^\m C^\n_{(1)}\sim F^{\m\n}_{(1)}\sim\n_0/R$. 
Thus the radiation gauge field has two different contributions. The 
first one, described by~(\ref{4.3}) and~(\ref{4.4}), corresponds to the usual 
result 
in electrodynamics. The second one can be obtained by solving the differential 
equations~(\ref{Cgauge}) and ~(\ref{4.10}) and 
corresponds to the interaction of the external field with the velocity fields
in $F^{\m\n}_{(0)}$. This additional contribution to the radiation fields 
stems from the non-linearity of the original system of coupled  field 
equations: 
for example, unlike in electrodynamics, an external electromagnetic wave 
interacts with 
the bulk scalar, gauge and graviton fields of the source;  this gives a  
contribution to 
the scalar, gauge field and gravitational radiation emitted. 

Although we were not able to find a solution of  the equations~(\ref{Cgauge}) 
and~(\ref{4.10}) in closed form, it is clear that these equations can be 
integrated. 
The solution should then be substituted in the stress-energy tensor for the 
gauge 
field given in eq.~(\ref{2.27}). The computations of the corresponding 
Poynting vector, 
of the differential and  total power radiated and of the Thomson scattering 
cross 
section are then similar to those given in section~\ref{scalarrad}, albeit 
more tedious.

\section{Gravitational Radiation from Accelerated Branes} 
\news
 
In this section we study the solution for the metric fluctuations. 
The linearised equation for $\bar{h}^{\m\n}$ in the case of the F-string 
was derived in section 2, with the result
\eqn
\label{5.1}
\arr{rcl}
\p^2\bar{h}^{\m\n}&=&
\displaystyle{-2\kappa^2M\int d\t U^\m(\t)U^\n(\t)\d^{(4)}(x-Y(\t))}\\\\
&&
\displaystyle{-2p\left(2\p^{(\m}\n_0\p^{\n)}\n-
\eta^{\m\n}\p\n_0\cdot\p\n\right)}\ .
\earr
\eeqn
In the case of  the D$p$-branes, there is an extra term of 
$-\frac{p}{2}\n_0(Y(\t))$ multiplying
the delta-function source; this  we  can ignore for reasons  explained in 
section 3 when
studying scalar field radiation. As in the case of  the gauge field,  we 
split the metric fluctuations as
$\bar{h}^{\m\n}=\bar{h}^{\m\n}_{(0)}+\bar{h}^{\m\n}_{(1)}$ and obtain the
equations
\eqn
\label{5.2}
\arr{rcl}
\p^2\bar{h}^{\m\n}_{(0)}&=&
\displaystyle{-2\kappa^2M\int d\t U^\m(\t)U^\n(\t)\d^{(4)}(x-Y(\t))}\ ,\\\\
\p^2\bar{h}^{\m\n}_{(1)}&=&
\displaystyle{2p\frac{A\kappa^2M}{4\pi}\frac{1}{[U\cdot (x-Y(\t))]^2}
\left(2\frac{(\p^{(\m}\n_0)(x-Y(\t))^{\n)}}{U\cdot (x-Y(\t))}
+2\p^{(\m}\n_0^{\n)}\right.}\\\\
&&\displaystyle{\left.\ \ \ \ \ \ \ \ \ \ \ \ \ \ \ \ \ \ \ \ \ \ \ \ \ \ 
\ \ \ \ \ \ \ \ \ \ \ 
-\eta^{\m\n}
\frac{(\p\n_0)\cdot(x-Y(\t))}{U\cdot (x-Y(\t))}\right)_{\t=\t_0}}\ .
\earr
\eeqn
The retarded solution for $\bar{h}^{\m\n}_{(0)}$
is
\eqn
\label{5.3}
\bar{h}^{\m\n}_{(0)}=-\frac{\kappa^2 M}{2\pi}
\left[\frac{U^\m(\t) U^\n(\t)}{U\cdot (x-Y(\t))}\right]_{\t=\t_0}\ .
\eeqn
Within our approximations, the only surviving terms in the equation for 
$\bar{h}^{\m\n}_{(1)}$ are the velocity fields of the  scalar $\n$. 
As in the case of the gauge field, the fluctuations $\bar{h}^{\m\n}_{(1)}$ are
expected to contribute to the observed radiation field.

To confirm this expectation consider the gauge condition 
$\p_\m\bar{h}^{\m\n}=0$.
For $\bar{h}^{\m\n}_{(0)}$ we have
\eqn
\label{5.4}
\p_\m\bar{h}^{\m\n}_{(0)}=-\frac{\kappa^2 M}{2\pi}
\left[\frac{\dot{U}^\n}{U\cdot (x-Y(\t))}\right]_{\t=\t_0}\ .
\eeqn
This has the correct $1/R$ fall off behaviour for a first derivative of the 
graviton
field to yield a non-vanishing energy flux at infinity. Using the equations 
of motion for the branes~(\ref{2.30}) and~(\ref{2.32}), we find that the gauge 
condition is satisfied if
\eqn
\label{5.5}
\p_\m\bar{h}^{\m\n}_{(1)}=-\frac{\kappa^2 M}{2\pi}
Ap\left[\frac{\p^\n\n_0(Y(\t))}{U\cdot (x-Y(\t))}\right]_{\t=\t_0}\ .
\eeqn
A hint for the above reasoning comes from consideration of the graviton 
equations 
of motion 
\eqn
\label{5.6}
\p^2\bar{h}^{\m\n}=-2\kappa^2\left(T^{\m\n}_{source}+T^{\m\n}_{\n}\right)\ .
\eeqn
The conservation equation $\p_\m T^{\m\n}=0$ (which implies that 
$\p^2\p_{\m}\bar{h}^{\m\n}=0$) is a necessary condition for 
$\p_{\m}\bar{h}^{\m\n}=0$ and,  after some algebra, gives the brane 
equations of  motion. 

To summarise: we found the solution for $\bar{h}^{\m\n}_{(0)}$
and the differential equations obeyed by $\bar{h}^{\m\n}_{(1)}$, which can be 
integrated (at least numerically). Both terms are expected to contribute to 
the 
observed radiation. To obtain the total radiated power and other quantities of 
interest, the solution for the metric fluctuations can be substituted in the 
formula 
for the stress-energy tensor of the gravitational field around flat space,
which is well-known in general relativity (see for example~\cite{SW}, 
chapter 7 
for a discussion).

\section{Discussion} 

In this paper, we have considered accelerated wrapped fundamental strings and 
D$p$-branes in the linear approximation to the classical limit of superstring 
theory. The results we obtained are thus valid provided the string coupling
$g_s$ is small and the magnitude of the acceleration $a\ll 1/\sqrt{\alpha '}$.
We showed that such objects generically produce scalar, gauge field 
(the appropriate NS-NS or R-R field with respect to which the brane is 
charged) 
and gravitational radiation. The field responsible for the acceleration  was 
taken to be an external scalar incident on the branes. The space-time in which 
the branes propagate was compactified to four dimensions on 
$T^p \times T^{(6-p)}$. 
We  neglected both the Ka\l u\.{z}a-Klein modes of a string or brane source 
around $T^{(6-p)}$ and the internal excitations of  a $p$-brane wrapped around 
$T^p$. It would be interesting to see if the inclusion of these modes leads to 
distinctive features of the emitted radiation. Alternatively, one can 
compactify 
on $T^p$ and obtain results similar to the four dimensional ones considered 
here for wrapped  $p$-branes in $D=(6-p)$ space-time dimensions. 

In the case of the bulk scalars, we were able to solve the linearised field 
equations and obtained a Larmor-type formula for the total radiated power. 
The angular distribution of the emitted radiation was studied. Finally, the 
cross section for scattering of the incident scalar wave was calculated; this 
gives an estimate for the classical radii of the branes. In the case of the 
gauge and gravitational fields, we were not able to solve the linearised field 
equations in closed form because of the coupling of the accelerating scalar 
field to the various fields produced by the moving source. However, we 
obtained differential equations which can be integrated; from the solutions 
to these equations, one can in principle derive the total power radiated in 
the form of gauge field and gravitational waves and also the cross sections 
for  scattering of the incident external perturbation. 
   
Our results may be viewed as a generalization to string theory and $p$-branes 
(including D-branes) of previous work by cosmologists on radiation from 
accelerated cosmic strings (for a review and references, see~\cite{VS}). 
The approach consists in solving the field equations of the 
Maxwell-Einstein-Higgs 
system, or a suitable supergravity generalization thereof, coupled to 
accelerated 
localised sources in the weak field approximation. The linearization of the 
field 
equations serves two purposes. First, it simplifies the otherwise formidable 
task 
of solving the system of coupled partial differential equations for the 
scalar, 
gauge field, gravitational and other fields in the problem. It also allows one 
to avoid discussing thorny issues which make their appearance beyond the 
linear 
approximation: for example, the full Einstein equations~(\ref{2.26}) in the 
presence of a sharply localised source involve products of $\delta$-functions 
which must be dealt with either using the theory of distributions or by 
attributing 
a small thickness to the source and taking an appropriate limit. The 
weak-field 
approximation is certainly valid for sources whose fields fall off 
sufficiently 
fast in the far away region, where the radiation is observed. The thin-source 
approximation is valid in the long wave-length limit of string theory, but 
will 
break down at high energies: there the dynamics is governed by the light open 
strings, which determine the effective thickness of the source.

Our  approach has also neglected a number of dynamical effects which it would 
be interesting to include in the analysis. For example, we have not discussed 
the back-reaction of the emitted radiation on the source: as energy and 
momentum 
are carried away by the emitted radiation,   the solution describing the 
source 
and its energy-momentum tensor will be modified. In addition, the emitted 
momentum 
will make the source recoil and accelerate. These effects should be of import 
in understanding the dynamics and fate of strings and branes accelerated in a 
curved background, like that of a non-extremal black hole. In the classical 
limit we have considered here, such effects  could be studied by viewing the 
actions~(\ref{2.12}) or~(\ref{2.14}) for the brane sources as probe 
actions~\cite{DPS},  
substituting  the pull-backs of the corresponding space-time metrics and 
solving 
the corresponding field equations in the weak-field approximation (which will 
be valid as long as the probe and the `heavy' object responsible for the 
acceleration are well separated). 

It would also be very interesting to understand accelerated branes and 
radiation 
effects from branes in the context of the AdS/CFT 
correspondence~\cite{Mald,GKP,Witten}. 
In~\cite{SD}, the interactions of  probe D3 branes in flat and in 
$AdS_5 \times S_5$ 
backgrounds were considered. In the latter case,  the effects of exchange of 
bulk 
supergravity modes carrying  momentum in the brane directions were related to 
certain 
terms involving derivatives of the field strength in the action of the dual 
gauge 
theory. It is plausible that such terms, along with 
other higher loop terms in the gauge theory, correspond to accelerated probes 
in the curved background. Although the acceleration terms in the gauge theory 
discussed in~\cite{SD} correspond in the bulk to NS-NS and R-R 2-form 
exchange, 
there should be additional terms corresponding to exchange of scalar and 
gravitational fields. It will be interesting to see whether the  emission 
of radiation in the bulk can be computed in the gauge theory from 
these terms.       

Another challenge to which our work points is of course to reproduce and 
obtain 
quantum corrections to results such as~(\ref{3.17}) and~(\ref{3.18})
on the radiation emitted by accelerated strings and $p$-branes directly 
from the underlying string theory. With current methods, this appears to be a 
rather daunting task: even for open F-strings ending on the D-branes 
(let alone NS fivebranes), 
the boundary conditions corresponding to an accelerated brane break the 
world-sheet 
conformal invariance. Clearly, some understanding of accelerated 
open string dynamics at the quantum level is required.    

\subsection*{Acknowledgements} 

We would like to thank Costas Bachas, Michael Green, Michael Gutperle,  
Daniel Holz, Hermann Nicolai, Lori Paniak and Malcolm Perry for helpful 
comments. 
The work of MSC was supported by FCT (Portugal) 
under the PRAXIS XXI programme and by  NSF grant PHY-9802484.

\end{document}